# Hearing Loss, Cognitive Load and Dementia: An Overview of Interrelation, Detection and Monitoring Challenges with Wearable Non-invasive Microwave Sensors

Usman Anwar, Tughrul Arslan, *Senior Member, IEEE,* and Amir Hussain, *Senior Member, IEEE*

*Abstract—* This paper provides an overview of hearing loss effects on neurological function and progressive diseases; and explores the role of cognitive load monitoring to detect dementia. It also investigates the prospects of utilizing hearing aid technology to reverse cognitive decline and delay the onset of dementia, for the old age population. The interrelation between hearing loss, cognitive load and dementia is discussed. Future considerations for improvement with respect to robust diagnosis, user centricity, device accuracy and privacy for wider clinical practice is also explored. The review concludes by discussing the future scope and potential of designing practical wearable microwave technologies and evaluating their use in smart care homes setting.

## I. Introduction

More than 5% of the world's population, or around 360 million people suffer from hearing loss. Hearing loss affects roughly 11 million people in the UK, making it the second most prevalent disability. Hearing loss increases sharply with age, and affects more than 40% of adults over the age of 50 in the UK, growing to more than 70% over the age of 70 [1]. Around 75% of the old population in care homes is disproportionately affected by some form of hearing loss. Healthy aging is linked to neurological and micro-vascular changes, which often means the onset of Age Related Hearing Loss (ARHL) and cognitive decline at the same time [2]. It is crucial to address hearing loss as early as possible after diagnosis, to diminish its adverse impact. There are several measures available for rehabilitation of people with hearing loss which includes use of hearing aids, middle ear and cochlear implants. Hearing aids can benefit 6.7 million individuals in the UK, but only 2 million use them. Unassisted hearing loss usually leads to severe health implications in older people which include social isolation, mobility loss and cognitive decline. In case of unassisted hearing loss, people with minor hearing loss are twice as likely as those without hearing loss to suffer from cognitive decline. People with moderate hearing loss are three times more likely to suffer from dementia, while those with severe hearing loss are five times more likely.

The rest of this paper is organized as follows: Section II reviews the effect of hearing loss on the brain. Section III and IV further explore links between hearing loss and cognitive decline and dementia respectively. Section V investigates potential benefits of cognitive load detection. Section VI discusses the possibility of utilizing hearing aid technology to reverse cognitive decline. This is followed by an overview of the use of wearable microwave sensors to detect cognitive load in Section VII. Finally, a number of challenges and opportunities relating to the practical design of wearable microwave sensors are outlined in Section VIII, including their potential application in future smart care homes.

## II. Effects Of Hearing Loss On Brain

Hearing loss or impairment usually occurs among older adults, and deteriorates healthy living in older adults [3]. Hearing loss causes a shift in cognitive resources from memory towards auditory processing, putting an undue strain on brain functions. In severe situations, this leads to cognitive decline and dementia.

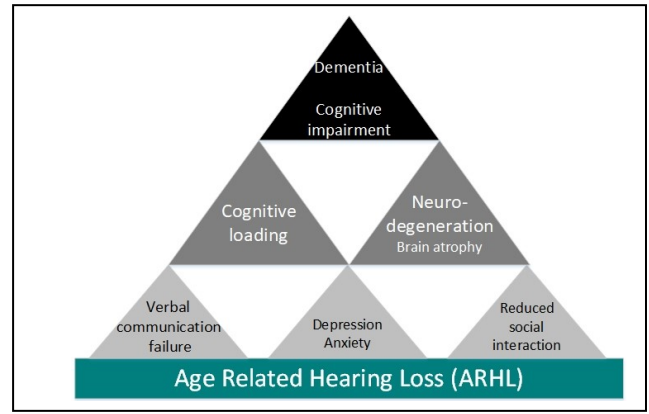

Figure 1. Association hierarchy of ARHL with dementia

Hearing loss usually coincides with other conditions including social isolation, anxiety, falls, depression, decreased quality of life, and memory and communication problems. ARHL has been linked to a quicker rate of cognitive decline. There are multiple theories associated with this. The first is related to cognitive load that happens as a result of untreated hearing loss. Hearing loss is more common in adults with dementia [3]. According to recent studies, dementia and Alzheimer's disease are more likely to occur in older adults with hearing loss [4]. The brain gets overworked by constant strain to comprehend audio and speech. More concentration is directed towards information deciphering and the brain becomes less focused on retaining that information. The second theory relates to social isolation and loneliness since untreated hearing loss causes a reduction in socializing. Social isolation usually results in less stimulation of the brain. Brain cells, particularly those that hear and process sound, can decrease due to lack of stimulation. This affects motor skills and leads to structural changes which can cause the brain to shrink. Long-term auditory deprivation can affect cognitive performance by lowering communication quality, which can lead to social isolation, depression, and dementia. [4]. Hearing loss is recurrently linked to cognitive deterioration in most studies. However, extensive research on the relationship between hearing impairment and Mild Cognitive Impairment (MCI) has only been undertaken in the last decade, and hearing impairment is linked to a higher risk of dementia and cognitive impairment [5].

Auditory deprivation results in social isolation and this leads to depression and decline in cognitive function. Long-term hearing loss shifts cognitive resources away from memory and toward auditory processing, putting undue burden on higher cortical functions that eventually leads to cognitive decline, complete association hierarchy of ARHL with dementia is given in Fig.1. Hearing loss can alter the auditory pathway and affects the auditory brain, resulting in dementia and cognitive decline [6]. Another hypothesis is that neurodegenerative diseases attack the auditory brain, resulting in peripheral and central hearing loss. These neuro-degenerative pathologies include canonical dementias such as lewy-body dementia, vascular dementia, Alzheimer's disease and frontotemporal dementia. Temporal, frontal, subcortical and parietal circuits that underpins auditory cognition are all compromised by these neurodegenerative pathologies [7]. The principal neurodegenerative causes of dementia in middle to later life are pathogenic protein distribution over large-scale cerebral networks and various forms of localized brain shrinkage. These are the key symptoms which can help to anticipate hearing impairments that accompany these specific dementia conditions.

III. COGNITIVE DECLINE AND HEARING LOSS

For a person affected with hearing loss, a constant attention, and effort to understand and decode speech must always be active. Hearing loss causes degraded auditory signals, auditory perceptual processing necessitates more cognitive resources, resulting in a cognitive shift away from other tasks and toward effortful listening. This phenomenon finally leads to cognitive reserve depletion and excessive cognitive load causes neurodegeneration with mild to severe brain changes [8]. Hearing impairment may be aggravated by this cognitive reserve depletion and this leads to degeneration in auditory perception. So the attention required to understand and comprehend speech is vital for individuals with hearing loss.

Hearing loss, according to the cognitive load hypothesis, results in degraded auditory signals. Increased cognitive resources required for auditory perceptual processing, and diversion from other cognitive tasks to effortful listening, eventually leads to cognitive reserve depletion. According to this proposition, an ageing brain suffers from a neurodegenerative process that causes both cognitive impairment and hearing loss. Another study established that hearing loss leads to cognitive deterioration which is either permanent or reversible with rehabilitation [4]. Hearing loss increases cognitive strain in patients with cognitive impairment, according to this study. Impaired perception may lead to deterioration in cognition and social seclusion, both of which can contribute to cognitive decline.

IV. DEMENTIA AND HEARING LOSS

Around 50 million people worldwide suffer from dementia, a figure that is expected to rise to 152 million by 2050. Furthermore, hearing loss affects over 465 million individuals, including one-third of those over the age of 65. According to the World Health Organization, untreated hearing loss costs the world economy $750 billion per year, while the current annual cost of dementia is around $1 trillion, with estimates that this could double by 2030. Hearing loss in middle age is supposed to contribute for 9% of dementia occurrences, and latest statistics suggest that dementia affects 47 million people worldwide. ARHL is usually caused by cochlear damage, while dementia occurs as result of cortical degeneration with an initial deterioration in the multimodal cortex. There are multiple studies which relate peripheral auditory decline to major cortical changes associated with dementia [8].

Recent studies linked severe dementia to intensive hearing loss. This means that hearing loss has a linear relationship with dementia, and that the risk of dementia increases multiple times with incremental severity of hearing loss [7]. Cognitive processing gradually slow with age, and at the age of 70, one-fifth of the population is estimated to have a considerable degree of cognitive loss [9]. According to this study, ARHL impacts severely on the brain and is assumed to be a primary reason for cognitive decline in older adults. ARHL has a negative influence on cognitive performance and raises the risk of dementia by adding more damage to existing brain impairments such as tau-tangles, amyloid-beta, brain atrophies and microvascular infarct [8]. Vascular brain anomalies, macro and microvascular infarction, aggravated as a result of ARHL, usually lead to vascular dementia and Alzheimer in most cases. Some studies have suggested a strong correlation between hearing loss and stroke. Central white matter pathways degradation and lower cortical volumes in the primary auditory cortex are also linked to peripheral hearing loss. Cochlear losses may result in cortical reorganization and degeneration of stria vascularis [10].

Hearing impairment may result in rapid brain shrinkage and decreased primary auditory cortex volumes in elderly adults, and ARHL is commonly associated with reduced brain volume. Grey matter volume reduction in the auditory cortex is suggested to be a probable cause of peripheral hearing loss in the elderly. [11]. Hearing ability and grey matter volume were found to have a significant linear relationship, and individual differences in hearing capacity predicted the amount of neuronal activation in temporal gyri, comprising the auditory cortex, brainstem, and thalamus.

Central auditory processing is the ability of the brain to understand sounds received by the cochlea. [1]. As a result, it is vulnerable to neurodegeneration, and data suggests that central auditory processing is impaired early in Alzheimer's disease and MCI. Central auditory impairment is a type of hearing loss, as measured by dichotic listening activities, and can be a precursor to Alzheimer's disease.

## V. Cognitive Load, Diagnosis And Repercussions

Listening effort to comprehend speech results in a higher utilization of cognitive resources for hearing impaired individuals. As cognitive reserves begin to diminish, this causes an undue cognitive strain on neurological functions. Cognitive load, if left untreated, may progress to dementia in later stage. Near-Infrared Spectroscopy (NIRS) and Magnetic Resonance Imaging (MRI) and) can be utilized for cerebral blood flow measurement but both technologies are expensive and require extensive medical supervision. EEG techniques use electrodes on the scalp to record the electrical response of the brain. Ionic motions in and around neurons generate these electrical signals during the activation and deactivation of neurons involved in cognitive tasks. The varying voltages in these electrical signals are measured by the EEG.

A positive linear relationship between cognitive load and blood pressure has been found in recent investigations [12], implying an increase in cognitive load will result in an increase in blood pressure. An increase in blood pressure can also cause cognitive load to increase [13]. As a result, different blood pressure levels can be used to assess the amount of cognitive load. Increased systolic blood pressure is easy to detect and is more likely to have an impact on the human body. Because blood pressure and metabolic rate are related in a positive linear way, the metabolic rate is also a major indicator to gauge the cognitive loading state.

## VI. Do Hearing Aids Help In Reversing Cognitive Decline?

Hearing aids or a cochlear implant can help with social and emotional functioning, communication, and cognitive performance, as well as improve overall quality of life. Detecting the severity of hearing impairment is vital to halt the disease progression, using prevention tools. Routine hearing care can be provided to the general public to protect cognitive function and reduce public health costs associated with MCI and dementia as they progress. Hearing examinations can be used clinically to assess the risk of MCI/dementia, and hearing aids can be used to delay dementia in older people with hearing impairment if the role of hearing impairment in cognitive decline and dementia is clarified. Currently available treatments include hearing aids, middle ear and cochlear implants that can delay the onset of cognitive decline. Recent studies in [14][15] have confirmed that the hearing aids are beneficial in the early stages of ARHL, and hearing aids can be helpful in rehabilitation of higher cortical functions. Use of hearing aid able to reverse central auditory system aging were reported in [2]. The study observed considerable improvements in cognition with hearing aid usage, and the results returned to baseline where hearing aid was not involved in human subjects.

## VII. Wearable Microwave Sensors For Cognitive Load Detection

In today's health sector, expensive magnetic resonance imaging (MRI), ultrasound, positron emission tomography (PET) and computerized tomography (CT) scan tests are widely in use for medical imaging applications. These techniques provide images of internal organs to medical practitioners for diagnosis of potential diseases. Results provided by these invasive methods differ in terms of resolution, operating method and implementation cost. Although these techniques are being widely used, they are not flexible and rely on ionizing radiations which may have adverse effects on longer exposure. In recent years, microwave sensing and imaging systems have gained a lot of attention. Microwave sensing and monitoring methods provide a more flexible, low cost, compact, non-invasive, low exposure and non-ionizing solution offering a potential replacement option for existing imaging equipment in the near future. Several methodologies have been reported recently for detection of tumors, brain stroke, Alzheimer's and breast cancer [16][17][18][19]. Microwave sensing technologies work on the basis of differences in electrical conductivity and permittivity between unhealthy and healthy regions of the human body. Electromagnetic (EM) waves are transmitted towards the target area of the body and reflected signals are collected back at the sensors for diagnosis of any potential disease in the target area. The received signals are scattered and reflected with varying intensity depending on the difference in electrical properties of that body area.

## VIII. Practical Design Of Wearable Microwave Sensors: Challenges And Opportunities

In contrast to traditional detecting technologies such as CT scan, X-ray, and MRI, microwave systems are considered safe for the human body and can be employed in medical applications. They provide a non-invasive, low-radiation technology that employs external scattering field to measure the target. The most critical factor in microwave detection is precision. The antenna sensor system is responsible for transmitting and receiving signals in microwave imaging systems, hence its performance has a direct impact on the imaging effect. It is critical to create an antenna that is miniaturized for microwave imaging systems and may include features such as ultra-wide bandwidth, high gain, and good directivity. This requires design optimization to ensure that the antenna receives accurate data, keeping in consideration the suitability of design material for biomedical applications. A number of challenges and opportunities relating to some of these design aspects are outlined below.

### A. Antenna miniaturization at low frequencies:

Designing antennas for wearable applications is challenging as it involves antenna miniaturization to make it compatible with the wearable device. Antenna miniaturization helps to reduce physical dimensions of the system while keeping the antenna functionality intact. For the miniaturization of antennas targeted at the elderly population, several design factors are required to be incorporated which include ease-of-use, non-intrusive, adaptable and portable interfaces. However, there are various challenges involved in the miniaturization of antennas operating at lower frequencies. The antenna can be made smaller by increasing the frequency and decreasing the wavelength. Further, the wearable body sensors application targeted at the brain require antenna sensors that operate at lower frequencies.

### B. Mutual coupling issue between antenna elements

To effectively scan target areas of the body, the wearable sensor system should contain multiple elements for extended coverage. Placement of antenna elements close to each other results in mutual coupling between them, which affects the overall gain, efficiency and creates distortion. Several techniques can be implemented to keep the spacing to an optimum level, introduction of Defected Ground Structure (DGS), Electromagnetic Bandgap (EBG) [20], parasitic elements and the use of metamaterials to enhance isolation between elements.

### C. Ensure cost-effective, portable and low power solutions

Wearable gadgets have recently gained much attention due to their wide range of applications in sports, navigation, military, medical and space industry. A major bottleneck in the optimization of electronic components for wearable applications is antenna design. These antennas must either be built into the human body, as wearables, or incorporated within the clothing. They must be low-cost, low-profile, portable, resilient, reliable and low-power, else they will be ineffective for these applications.

### D. Mitigation of near-body effects on wearable sensor systems and Specific Absorption Rate (SAR)

Microwave systems designed for wearable applications usually operate within close proximity to the human body. This not only has an impact on the performance of sensors but can also cause damage to human health through electromagnetic radiations. Radiation intensity is a major challenge to consider while designing wearable antennas. SAR is a criterion for determining if an antenna is safe for the human body, and it refers to the ratio of radiation generated by the antenna to the amount of absorption. There are two standards in place, IEEE standard allows maximum absorption of 1.6W/Kg for 1g and International Commission for Non-Ionizing Radiation Protection permits maximum of 2W/Kg per 10g [21].

### E. Smart Care Home Applications

Ongoing future work, as part of the COG-MHEAR project [22], aims to develop and evaluate the use of emerging wearable non-invasive sensing technologies [23] for the detection of cognitive load within a hearing aid device that could be used by people with hearing loss in smart care home settings in order to enhance their quality of life and provide better care services.

## IX. CONCLUSION

Non-invasive wearable sensors are being developed for conditions such as dementia and neurodegeneration detection. The integration of these sensors with new sensors being developed for cognitive load monitoring within a hearing aid device, could lead to more effective procedures for the detection and treatment of more complex conditions arising in care settings. In future, these wearable devices can be linked to the cloud and the data obtained can be post-processed with cloud analytics and machine learning for better diagnostics and care services.